\documentstyle[twocolumn,aps,prl,epsf,floats]{revtex}
 \def\half{\mbox{$1\over 2$}}

 \def\beq{\begin{equation}}
 \def\eeq{\end{equation}}
 \def\beqa{\begin{eqnarray}}
 \def\eeqa{\end{eqnarray}}
 
 \def\LP{\left(}
 \def\RP{\vphantom{\half} \right)}
 \def\vph{\vphantom{\hat E}}
 \def\leftp{\left(\vph\!\!\right.}
 \def\rightp{\left.\vph\!\!\right)}
 
 \def\real{\Re {\rm e}}

 \def\ltsim{\stackrel{<}{\sim}}
 \gdef\aver#1{\left\langle #1 \right\rangle}
 \gdef\s#1{\! #1 \!}
 \gdef\l#1{\> #1 \>}
 \gdef\Eq#1{Eq.~(\ref{#1})}

\begin{document}
\draft


\title{
Nonequilibrium Josephson--like effects
in wide mesoscopic S--N--S junctions}

\author{Nathan Argaman\cite{curadd}}
\address{ Institute for Theoretical Physics,
University of California, Santa Barbara, CA 93106, USA }


\address{\parbox{14.5cm}{\rm\small
\medskip
Mesoscopic superconducting--normal-metal--superconducting (S--N--S)
junctions with a large separation between the superconducting electrodes (i.e.\ wide junctions) exhibit non\-equilibrium supercurrents, even at temperatures for which the equilibrium Josephson effect is exponentially small.  The second harmonic of the Josephson frequency dominates these
currents, as observed in recent experiments.  A simple description of these effects, in the spirit of the Resistively--Shunted--Junction model, is suggested here.  It is used to calculate dc $I$--$V$ characteristics, and to examine the effects of various types of noise and of external microwave radiation (Shapiro steps).  It is found that the nonequilibrium supercurrents are excited when the junction is driven by a dc bias or an ac bias, or even by external noise.  In the case of junctions which are also long in the direction perpendicular to the current flow, thermodynamic phase fluctuations (thermal noise) alone can drive the quasiparticles out of local equilibrium.  Magnetic flux is then predicted to be trapped in units of $\Phi_0/2 = hc/4e$.
}}

\address{\parbox{14.5cm}{\bigskip \rm \small
}
}
\maketitle

\ifpreprintsty \vfill\newpage \fi

\section{Introduction}

Wide mesoscopic superconducting--normal-metal--superconducting (S--N--S)
junctions exhibit a spectrum of low--lying electronic ``Andreev bound
states'' \cite{minigap}, which depends strongly on the
phase--difference $\phi$ between the two superconductors \cite{eg}.  This $\phi$--dependence persists when the temperature $T$ is raised, even if the normal--metal coherence length, $\xi_{\mathrm N}$, becomes much smaller than the distance $L_x$ between the two superconducting electrodes.  In this high--temperature regime the equilibrium Josephson coupling is exponentially weak, but the effects of quantum--mechanical coherence of the electrons decay only as $1/T$ (Ref.~\cite{Courtois}).  As the spectrum depends on $\phi$, and hence on time $t$ (the derivative $d\phi/dt$ is
proportional to the voltage $V$), non--equilibrium (NEQ) effects
occur, including supercurrents which are the topic of this work.

\begin{figure}[t]
\epsfxsize=\hsize
\epsffile{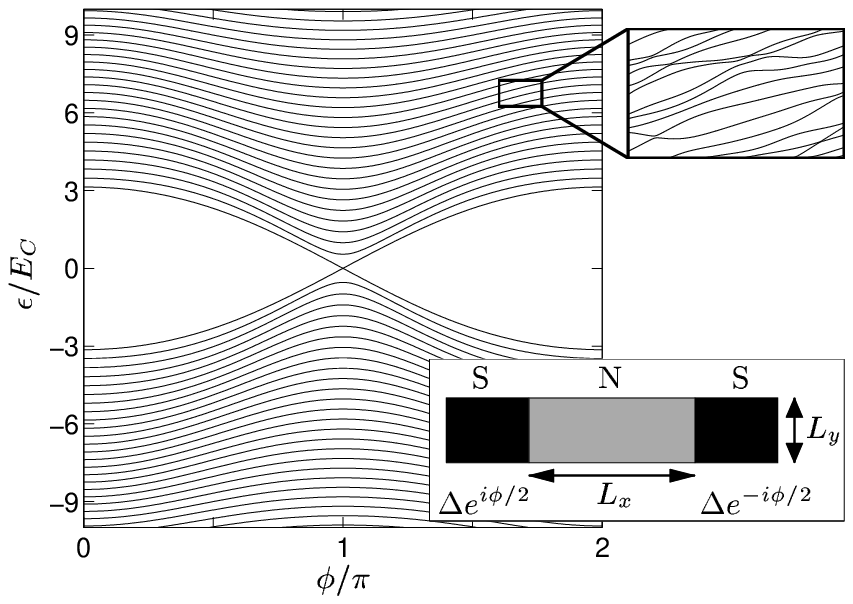}
\refstepcounter{figure}
\label{fig_dos}
FIG. \ref{fig_dos}: 
Energies of Andreev bound states $E_n$, equally spaced in
$n$, for a diffusive S--N--S junction (inset,
$E_{\mathrm C} \s= \hbar D/L_x^2 \ll \Delta$), plotted as a 
function of the superconducting phase--difference $\phi$.  
A phase--dependent mini--gap is proximity--induced at the 
Fermi level ($\epsilon \s= 0$).
Blowup (schematic): the avoided crossings of individual disorder
realizations, which are not resolved here as the spectrum is dense.
\end{figure}

NEQ proximity effects have been studied intensively for more than 20
years \cite{Tinkham}, but experiments in the mesoscopic regime have only
recently become possible.  The latter include Refs.~\cite{FEL,Bfield,dfrntl,Konrad} which displayed intriguing frequency--doubling effects and thus motivated the development of the theory presented here (an  interpretation in terms of the frequency--doubling effects predicted in the early eighties by Spivak and coworkers \cite{prev} was initially proposed, but had to be rejected \cite{Jack}).
Theoretically, NEQ phenomena have been studied in both clean
\cite{Zaikin,Wurzburg,Shumeiko} and dirty
\cite{Lempitskii,Averin,Spivak,Volkov,sample} limits. 
The approach taken here, following e.g.\ Ref.~\cite{Bagwell}, is to consider the junction as a mesoscopic system, or ``electron box''.  The special superconducting ``walls'' of this box impose a time--dependent boundary condition --- the external phase--difference $\phi$.  Once the $\phi$--dependence of the spectrum of the system is known, the supercurrents are simply given by the ``slopes'' of the occupied levels, $dE_n/d\phi$.  In fact, it is shown that, for evaluating the NEQ supercurrent, all of the different levels may be replaced by a single representative state, at a relatively low cost in terms of loss of accuracy ($\sim$15\%).
This results in a particularly transparent description of short junctions, $L_y \sim L_x$ (see inset in Fig.~\ref{fig_dos}),
which may be used in a wide variety of applications, in analogy to the well--known resistively shunted junction (RSJ) model.  The simplicity of the model allows us to consider for the first time NEQ--dominated long junctions, $L_y \gg L_x$.  Although a different language is used, the agreement with the quasiclassical Green's functions approach used in Refs.~\cite{Zaikin,Lempitskii,Spivak} is complete, as we will show explicitly.

We will assume that the normal--metal part of the junction is mesoscopic in the sense that the dephasing length $l_\varphi$ is much larger than $L_x$, but that the dephasing time $\tau_\varphi$ is sufficiently short to prevent the distinction between individual quantum levels, $\hbar/\tau_\varphi \gg \delta$, where $\delta$ is the single--particle level spacing (in the present context, a system with resolved levels, $\hbar/\tau_\varphi \ltsim \delta$, would best be called microscopic).  This implies that we are discussing a two-- or three--dimensional system, in which the number of transverse channels 
$N_\perp \sim \hbar v_{\mathrm F} / L_x \delta$ is very large ($v_{\mathrm F}$ is the Fermi velocity).  The effects of electron--electron and electron--phonon interactions in the normal--metal part will be ignored \cite{ee}, except for their inclusion in the dephasing rate
$1/\tau_\varphi$ and the energy--relaxation rate $1/\tau_{\mathrm E}$, which are treated as phenomelogical parameters.  However, the BCS interaction is assumed to dominate in the superconducting electrodes, so that they form good reservoirs for Cooper pairs, with well--defined phases (or at least phase difference, $\phi$) and a large gap $\Delta$ for single--particle excitations.

We focus on the ``high--temperature regime'', 
$E_{\mathrm C} \ll T \ll \Delta$, where $E_{\mathrm C}$ is
the Thouless energy or correlation length of the spectrum (the temperature $T$ is in energy units, $k_{\mathrm B}=1$).  The first inequality here is equivalent to $\xi_{\mathrm N} \ll L_x$, and implies that the equilibrium Josephson effect is exponentially supressed.  The second implies that single--particle excitations are ``bound'' to the normal--metal region and do not occur in the superconducting electrodes --- any electron or hole which approaches the N--S interface is either reflected normally or Andreev reflected back into the normal--metal region (Andreev reflections are coherent processes by which an electron can add a Cooper pair to the superconductor while evolving into a hole, or vice--versa, a hole can remove a Cooper pair and evolve into an electron; these processes convert current carried by Cooper pairs to current carried by single--particle excitations; re--diagonalizing the Hamiltonian in their presence, i.e.\ solving the Bogoliubov --- de Gennes equations, gives the Andreev bound states, which form the single--particle excitation spectrum of the system \cite{Tinkham}).  The Thouless energy $E_{\mathrm C}$ is defined as $\hbar$ over the time it takes an electron to cross the junction.  In the dirty limit, $l_{\mathrm el} \ll L_x$ with $l_{\mathrm el}$ the elastic mean--free-path, the Thouless energy is given by 
$E_{\mathrm C} = \hbar D / L_x^2$, with $D$ the diffusion constant.  In this case, the normal--metal coherence length is $\xi_{\mathrm N} = \sqrt{\hbar D/ 2\pi T}$.   We will also discuss the clean or quasiballistic case, $l_{\mathrm el} \gg L_x$, where $E_{\mathrm C} = 
\hbar v_{\mathrm F} / L_x$ and $\xi_{\mathrm N} = \hbar v_F/ 2\pi T$.  However, we will not discuss the super--clean limit, $l_{\mathrm el} \gg N_\perp L_x$, for which $\hbar/\tau_{\mathrm el} \ltsim \delta$ where $\tau_{\mathrm el}=l_{\mathrm el}/v_{\mathrm F}$ --- in that case disorder is irrelevant and the spectrum may become seperable.  Our discussion will also be limited to slow energy--relaxation rates, $\hbar/\tau_{\mathrm E} \ll E_{\mathrm C}$.  This inequality is already implied by our previous assumptions, because  $\tau_{\mathrm E} > \tau_\varphi$ and the dephasing length is large, $l_\varphi \gg L_x$.

The layout of this paper is as follows.  The model is developed in Sec.~II, including simple applications to NEQ ac and dc Josephson--like effects.  Further applications, including long junctions, are analyzed in Sec.~III.  The model is adjusted to describe a clean, quasiballistic system in Sec.~IV, and conclusions are briefly discussed in Sec.~V.  The consistency of the present treatment with the standard but more opaque Green's function approach is demonstrated in the appendix.

\section{Equilibrium and nonequilibrium supercurrents}

A mesoscopic junction may be characterized by a density of states, $\nu(\epsilon,\phi)$, which depends on the excitation energy $\epsilon$ (measured from the Fermi surface) and the superconducting phase--difference $\phi$.  In this section we will take this spectrum as input, and calculate the resulting supercurrents.

Let us begin by displaying the spectrum for a simple diffusive junction, for specificity.
The disorder--averaged density of states, $\nu(\epsilon,\phi)$, can be
calculated from the Usadel equations \cite{minigap,Argaman}.  It is
convenient to define the energy $E_n(\phi)$ of the $n$th level: $n =
\int_0^{E_n} \nu(\epsilon,\phi) \; d\epsilon$, with $E_{-n}=-E_n$, see Fig.~\ref{fig_dos} (negative excitation energies are included here for clarity of presentation; they represent the ``hole branch'' of the excitation spectrum).
The calculation assumes time--reversal symmetry, with the pair
potential vanishing in N (no electron interactions), and perfect N--S
interfaces (i.e.\ no potential barriers or Fermi velocity mismatch).
A mini--gap is induced in N by the proximity of S.  Its size is
$E_{\mathrm g} \simeq 3.1 E_{\mathrm C}$ at $\phi=0$, and it closes at 
$\phi = \pi$ and reopens periodically \cite{minigap}.  We are considering only the disorder--averaged spectrum, because the fluctuations in $E_{\mathrm g}$ and $E_n$ are only $\sim \delta$, and are associated with currents that are smaller than the ones considered here by a factor of $\delta/E_{\mathrm C}$.  This, the $\phi$--periodicity, and the scale $E_{\mathrm C}$ of the structure in the spectrum near the Fermi level, are generic to all junctions to be considered here.  Other features, such as the existence of a strict gap, are not generic and will not be relied upon --- for example, due to the great sensitivity of the spectrum to various types of symmetry breaking, adding spin--flip scattering with 
$\hbar/\tau_{\rm sf} \sim E_{\mathrm C}$ can close the mini--gap. Another concrete example --- a clean junction --- will be considered in Sec.~IV.

The current $I$ in the junction is related to the spectrum.
In the presence of a voltage $V = (\hbar/2e)(d\phi/dt)$ (the Josephson
relation), the energy of the electronic system, $\sum_n E_n f_n$,
is time--dependent \cite{condens}.  Dividing the power by the
voltage, one finds $I = I_{\mathrm S} \s+ I_{\mathrm N}$ with
\beq \label{Isum}
I_{\mathrm S} = {2e \over \hbar} \! \sum_{n=-\infty}^\infty \! 
{dE_n \over d\phi} f_n \quad ; \quad
I_{\mathrm N} =   \! \sum_{n=-\infty}^\infty \! {E_n \over V} 
\leftp {df_n \over dt} \rightp_V  \> ,
\eeq
where $I_{\mathrm S}$ is independent of $V$, and will be called the
supercurrent, and $I_{\mathrm N}$ is proportional to $V$ (dissipative), and will be called the normal current.  As is by now well--known (see e.g.\ Ref.~\cite{Bagwell}), the same distinction between a dissipative current and a persistent current can be made in a normal--metal mesoscopic ring with a non--trivial boundary condition $\phi$ imposed by an Aharonov--Bohm flux.  In that case the contributions of the different levels $E_n(\phi)$ to the persistent current $I_{\mathrm S}$ do not add up in a systematic manner, the resulting currents are much smaller than in S--N--S junctions, and in a disorder--averaged model of the type considered here, would vanish altogether.

As we are considering a system with a dense spectrum, the occupations $f_n$ do not evolve adiabatically (in contrast to Ref.~\cite{Averin}), but diffuse in energy \cite{Wilkinson} with a coefficient 
$D_{\mathrm E}(\epsilon,\phi) \sim G_{\mathrm N} V^2 \delta$, where $G_{\mathrm N} \sim (e^2/h) E_{\mathrm C}/\delta$ is the conductance of the normal--metal part in the absence of proximity effects.  This can be argued qualitatively by observing that in a time $t$, a quasiparticle crosses the junction $\sim t E_{\mathrm C}/\hbar$ times, and undergoes Andreev reflections which shift its energy by $\pm eV$ every time it encounters the N--S interface, implying difusion with 
$D_{\mathrm E} \sim (eV)^2 D/L_x^2$.  For $T \gg E_{\mathrm C}$ regions of weak proximity effects, $|\epsilon| > E_{\mathrm C}$, dominate in the normal current, and we will thus set $I_{\mathrm N} = G_{\mathrm N} V$.

While the occupations $f_n$ follow the slope of the $E_n(\phi)$ curves and diffuse in energy, they are also subject to electron--electron and electron--phonon interactions.  We use the relaxation time approximation \cite{imbal},
\beq \label{relax}
{df_n \over dt} \l\simeq 
       -{1 \over \tau_{\mathrm E}} \LP f_n \s- f_{\rm eq}(E_n) \RP  \; ,
\eeq
where $f_{\rm eq}(\epsilon) = 1/ \LP 1+\exp(\epsilon/T) \RP$, and 
consider only small voltages, where the term 
$D_{\mathrm E} (\partial^2 f/\partial \epsilon^2)$ which was dropped from the right hand side of \Eq{relax} can indeed be neglected,
and heating is unimportant ($T$ is time--independent).  This implies 
$\tau_{\mathrm E} D_{\mathrm E} \ll E_{\mathrm C}^2$, or 
$eV \ll \sqrt{E_{\mathrm C} \hbar / \tau_{\mathrm E}}$, which
is more stringent than the condition $eV \ll E_{\mathrm C}$ required
for the spectral density itself to be a meaningful quantity (at higher
voltages the spatial dependencies of the relevant Green's functions must also be taken into account).

In equilibrium, $I_{\mathrm S}  =  \int_{-\infty}^\infty d\epsilon \, 
j(\epsilon,\phi) / \! \LP 1 \s+ \exp(\epsilon/T) \RP = 
I_{\rm eq}(\phi)$,
where $j(\epsilon)$ is the ``Josephson current density'', 
$j(E_n,\phi) \propto \nu(E_n,\phi) (dE_n/d\phi)$, which is analytic in
the upper half of the complex $\epsilon$ plane (a positive
imaginary part of $\epsilon$ corresponds to a dephasing rate,
$\hbar/\tau_\varphi$, which would smooth out any singularities).
By contour integration one finds the well--known Matsubara sum:
$I_{\rm eq} = 2 \pi i T \sum_{m=1}^\infty j(i\omega_m)$, where 
$\omega_m = (2m \s- 1) \pi T$.  At high temperatures, even the smallest Matsubara frequency is $\gg E_{\mathrm C}$ and has a decay time $\hbar/\omega_1 < L_x^2/D$, yielding an exponentially small
$j(i\omega_m)$.  For physical quantities of this form,
thermal averaging is thus ``equivalent'' to dephasing.

In NEQ situations contour integration cannot be used, and the currents
are not exponentially small.  The large--$T$ behavior of $I_{\rm eq}$
implies oscillations of $j(\epsilon)$ such that $\int_0^\infty
d\epsilon \, \epsilon^k \, j(\epsilon,\phi) = 0$ for any odd power $k$, as can be seen by expanding $f_{\mathrm eq}$ in powers of $1/T$.
In the diffusive case, $j(i\omega)$ decays exponentially with
$\sqrt{\omega/E_{\mathrm C}}$ on the imaginary axis, and correspondingly
$j(\epsilon)$ oscillates and decays rapidly with
$\sqrt{\epsilon/E_{\mathrm C}}$ (the curves of Fig.~\ref{fig_dos} change their character repeatedly at higher $\epsilon$, occasionally having shallow maxima at $\phi = \pi$ rather than minima).  As can be seen from
\Eq{relax}, the deviations of $f_n$ from equilibrium change sign in
rhyme with these oscillations (and decay with $\epsilon$), and thus
the integrand of the NEQ part of the supercurrent, 
$2\int_0^\infty d\epsilon \, j \, (f_n \s- f_{\rm eq})$,
does not change sign and cannot be affected by any cancellations.

\begin{figure}[tb]
\epsfxsize=\hsize 
\epsffile{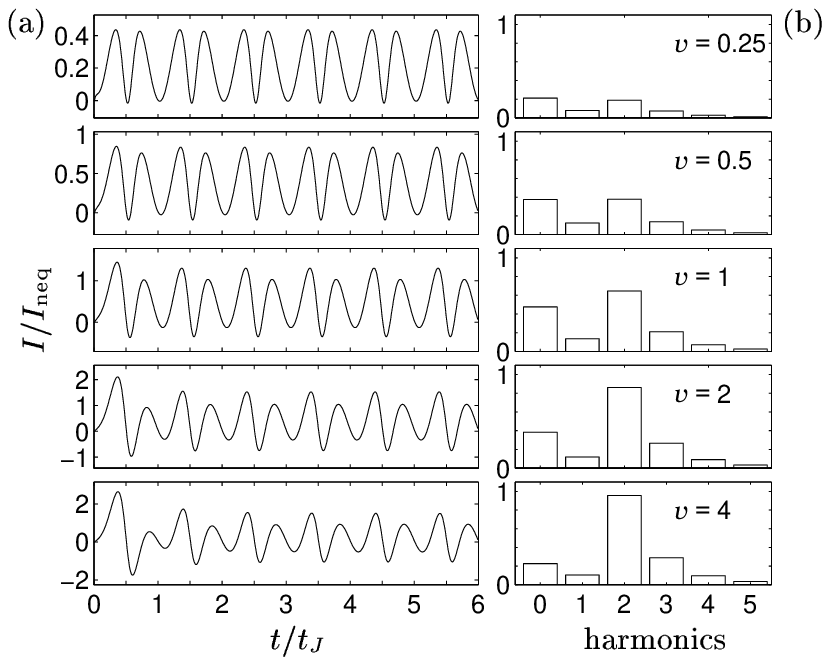}
\vspace*{1.5mm}
\refstepcounter{figure}
\label{fig_ac}
FIG. \ref{fig_ac}:
Supercurrents for a dc voltage bias: 
(a) $I_{\mathrm S}$ vs.\ $t$, and (b) amplitudes $I_k$ of its harmonic 
decomposition at $t \gg \tau_{\mathrm E}$ (with $k \geq 0$; phases not shown),
for five dc voltages labeled by $v = (2e/\hbar) V \tau_{\mathrm E}$.   
The Josephson period is $t_{\mathrm J} = h/2eV$ (for $I_{\text{neq}}$ see below). 
\end{figure}

When a dc voltage is applied to the junction, ${d\phi / dt} =
2eV/\hbar = {\rm const.}$, Eqs.~(\ref{Isum}) and (\ref{relax}) give
rise to ac supercurrents, see Fig.~\ref{fig_ac}.
Here $T \gg E_{\mathrm C}$ was used to take $f_{\rm eq}(\epsilon) = 
\half-\epsilon/4T$, and $I_{\rm eq}=0$.  The second harmonic dominates because $I_{\mathrm S}$ involves products of ${dE_n / d\phi}$ and 
$(f_n \s- f_{\rm eq})$, and both factors oscillate at the Josephson
frequency.  This ``frequency doubling'' is an essential feature in
identifying these effects experimentally \cite{Konrad}.

The limits of small and large voltages relative to the inelastic scatterring rate may be analyzed analytically \cite{Tinkham,Spivak}.  For small $(2e/\hbar)V \tau_{\mathrm E}$, the deviations of the occupations from equilibrium are linear in the voltage, 
$f_n \simeq f_{\mathrm eq} - 
\tau_{\mathrm E} (2e/\hbar) V (df_{\mathrm eq}/d\phi)$, where 
$df_{\mathrm eq}/d\phi \simeq (-1/4T)(dE_n/d\phi)$ for $T \gg E_n$.  The NEQ supercurrent is then disipative --- the energy is dissipated by the fast ralaxation $\tau_{\mathrm E}$ and not stored in the system. 
From \Eq{Isum}, one finds that
\beq \label{smallV}
G(\phi) = G_{\mathrm N} +
\LP {2e \over \hbar} \RP^2 {\tau_{\mathrm E} \over 4T} 
\sum_{n=-\infty}^\infty \LP {dE_n \over d\phi} \RP^2  \; ,
\eeq
with $I = G(\phi) V$.  Despite the fact that $I_{\mathrm S}$ is dissipative in this limit, we still use \Eq{Isum} to distinguish the component of the current we call ``normal'' from the ``supercurrent'' or ``persistent current'' which remains finite even when the voltage vanishes and equilibrium is restored, provided the temperature is sufficiently small.

In the opposite limit of large dc voltages, $(2e/\hbar)V \tau_{\mathrm E} \gg 1$, the individual occupation probabilities $f_n$ relax to the time--average of their equilibrium value,
\beq \label{largeV}
f_n \simeq {1 \over 2\pi} \int_0^{2\pi} {\mathrm d}\phi \; 
f_{\mathrm eq}\LP E_n(\phi) \RP 
\simeq {1\over 2} - {\overline{E_n} \over 4T}  \; ,
\eeq
where $\overline{E_n}$ is the phase--averaged value of $E_n(\phi)$.  The supercurrent is then obtained directly from \Eq{Isum}, and is ``purely reactive''.  In fact, the rate of dissipation in this limit is proportional to $1/\tau_{\mathrm E}$ and may be calculated from \Eq{relax}.  As it is independent of the voltage, the dc component of the current dacays as $1/V$ at large voltages (within the domain of validity of the model) see Fig.~\ref{fig_ac}.

The presence of the sum over $n$ (the energy integration) in \Eq{Isum} makes calculations according to this model relatively cumbersome.  In fact, it may often be superfluous, because the relaxation--time approximation used here already ignores possible $\epsilon$--dependencies, e.g. in $\tau_{\mathrm E}$.  One may also note that only the shape of the $E_n(\phi)$ curves is significant --- a simple shift in energy does not affect their contribution.  We will thus adopt a much simpler scheme, in which only the first $N$ states above (or below) the Fermi level are considered to have a significant $\phi$--dependence, and are all represented by a single curve, $E_{\mathrm rep}(\phi)$:
\beq \label{rep}
I_{\mathrm S}  \l\simeq  I_{\rm eq}(\phi) + 2 {2e \over \hbar} N 
 {dE_{\rm rep} \over d\phi} \, \LP f - f_{\rm eq}(E_{\rm rep}) \RP  \; ,
\eeq
where $f = {1 \over N} \sum_{n=1}^N f_n$ is the average occupation of the represented band of $N$ levels \cite{strict}.  For the system of Fig.~\ref{fig_dos}, 
$E_{\rm rep} =  3.4 E_{\mathrm C} \sqrt{1 \s+ 0.7 \cos(\phi)}$ and 
$N = 10 E_{\mathrm C}/\delta$ (including a factor of 2 for spin) 
reproduce the result of the full sum over $n$ very well ---
deviations are less than 15\%, for 85\% of the harmonic amplitudes
shown in Fig.~\ref{fig_ac} (the results plotted are from the simpler
model).

\begin{figure}[t]
\epsfxsize=\hsize
\epsffile{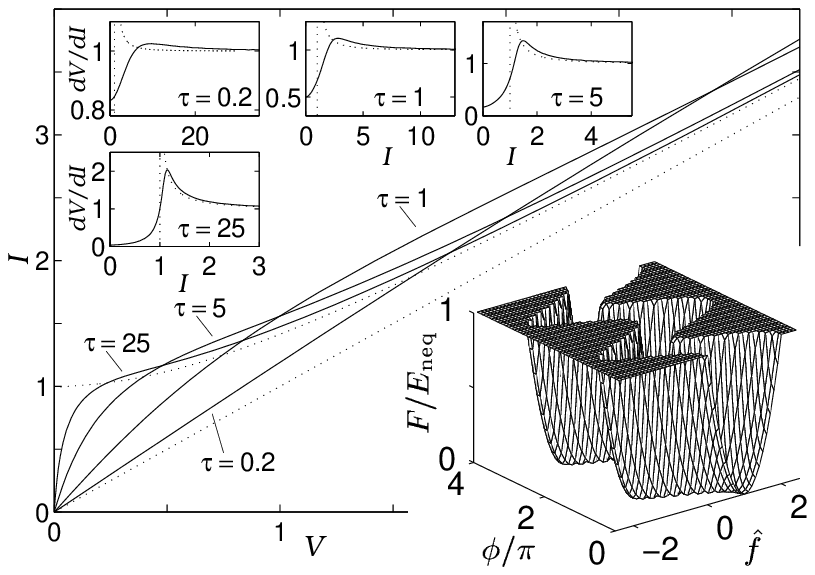}
\refstepcounter{figure}
\label{fig_IV}
FIG. \ref{fig_IV}:
The $I$---$V$ curves and differential resistances (upper insets) for a
dc current bias in the NEQ model, for four values of the relaxation time
$\tau_{\mathrm E}$; compared to the RSJ model with $I_{\mathrm c} = I_{\rm neq}$, and to $I = G_{\mathrm N} V$ (dotted lines). 
Lower inset: The effective free--energy function $F(\phi,\hat f)$, for
$I=0$, is a ``winding valley'' with minimum at $\hat f = \hat E(\phi)$
and a parabolic cross--section.
\end{figure}

Often, $I(t)$ rather than $V(t)$ is known.  In considering the conventional Josephson effect in such situations, it is customary to use the resistively shunted junction (RSJ) model \cite{Likharev}, which describes the dynamics of $\phi$ through 
$(\hbar/2e)(d\phi/dt) = (I-I_{\mathrm S})/G_{\mathrm N}$, with 
$I_{\mathrm S} = I_{\mathrm c} \sin \phi$ as appropriate for the equilibrium case.  This corresponds to overdamped motion of a ``representative point'' $\phi$ in a ``tilted washboard
potential'', $F(\phi) = - (\hbar/2e) (I_{\mathrm c} \cos \phi+I\phi)$.  To generalize this to NEQ situations where the occupation $f$ needs to be followed, one replaces the expression for $I_{\mathrm S}$ by \Eq{rep}, and describes the dynamics of $f$ by \Eq{relax}.  This is associated with the free--energy landscape
\beq \label{mod}
F(\phi,\hat f)  \l=  E_{\rm neq} (\hat E-\hat f)^2 - (\hbar/2e) I \phi 
\eeq
(see inset in Fig.~\ref{fig_IV}), where the overdamped motion is given by
\beq \label{dervs}
{\hbar \over 2e} {d\phi \over dt}  = 
   - {1 \over G_{\mathrm N}} \, 
     {2e \over \hbar} {\partial F \over \partial \phi} 
\quad ; \quad
{d\hat f \over dt}  =
   - {1 \over 2\tau_{\mathrm E} E_{\rm neq}} {\partial F \over \partial \hat f}  \; .
\eeq
Here $\hat E$ and $\hat f$ are rescaled quantities, 
$E_{\rm rep}(\phi) = A \hat E +B$ and 
$f = \half - (A \hat f +B)/4T$, with the requirement 
$-1 \leq \hat E \leq 1$ fixing $A$ and $B$ (in the present case,  
$A\simeq 1.3 E_{\mathrm C}$).  The current scale is given by
$I_{\rm neq} = (2e/\hbar) N A^2/4T$, and the energy scale is 
$E_{\rm neq} = (\hbar/2e)I_{\rm neq} 
\simeq 4.1 E_{\mathrm C}^3/\delta T$,
only a factor of $E_{\mathrm C} / 3.5 T$ smaller than the $T=0$
Josephson coupling energy, 
$E_{\mathrm J}(0) \simeq 14 E_{\mathrm C}^2/\delta$.

The dc $I$---$V$ curves of this model, obtained by taking 
$I(t) = {\rm const.}$, are displayed in Fig.~\ref{fig_IV},
using units of $I_{\rm neq}$, $V_{\rm neq}=I_{\rm neq}/G_{\mathrm N}$ 
and $\tau_{\rm neq} = \hbar/2eV_{\rm neq}$ for current, voltage and
time.  Note that $eV_{\mathrm neq} \sim E_{\mathrm C}^2/T$ whereas \Eq{relax} is restricted to $eV \ll \sqrt{E_{\mathrm C} \hbar/\tau_{\mathrm E}}$; thus the model may lose its validity before or after one reaches a voltage of one $V_{\mathrm neq}$ unit, depending on which of the large parameters $T/E_{\mathrm C}$ or $\sqrt{\tau_{\mathrm E} \hbar/E_{\mathrm C}}$ is larger.

The results for large $\tau_{\mathrm E}$ lie remarkably close to the dc 
$I$---$V$ curves of the RSJ model \cite{fzbc}, except for the
finite zero--bias conductance, equal to
$G_{\mathrm N} + \tau_{\mathrm E} (2e/\hbar) I_{\rm neq}$, or $1+\tau$ in the units used in the figure.  Indeed, for small currents and
voltages one may follow the analysis of \Eq{smallV}, and find that
$\hat f \simeq \hat E - \tau_{\mathrm E} (d\hat E/d\phi)(d\phi/dt)$,
and $G(\phi) = G_{\mathrm N} + 
(2e^2 N \tau_{\mathrm E}/\hbar^2 T) (dE_{\rm rep}/d\phi)^2$;
when integrated over a full period, our specific choice of form for $E_{\rm rep}(\phi)$ leads to the simple $1+\tau$ result.

\section{Applications, long junctions}

As in the RSJ model \cite{Likharev}, one may use Eqs.~(\ref{mod}) and (\ref{dervs}) to study a variety of more complicated physical situations, of which three will be considered here: the response of the junction to rf fields, the effects of noise, and the case of long junctions.

\begin{figure}[t]
\epsfxsize=\hsize 
\epsffile{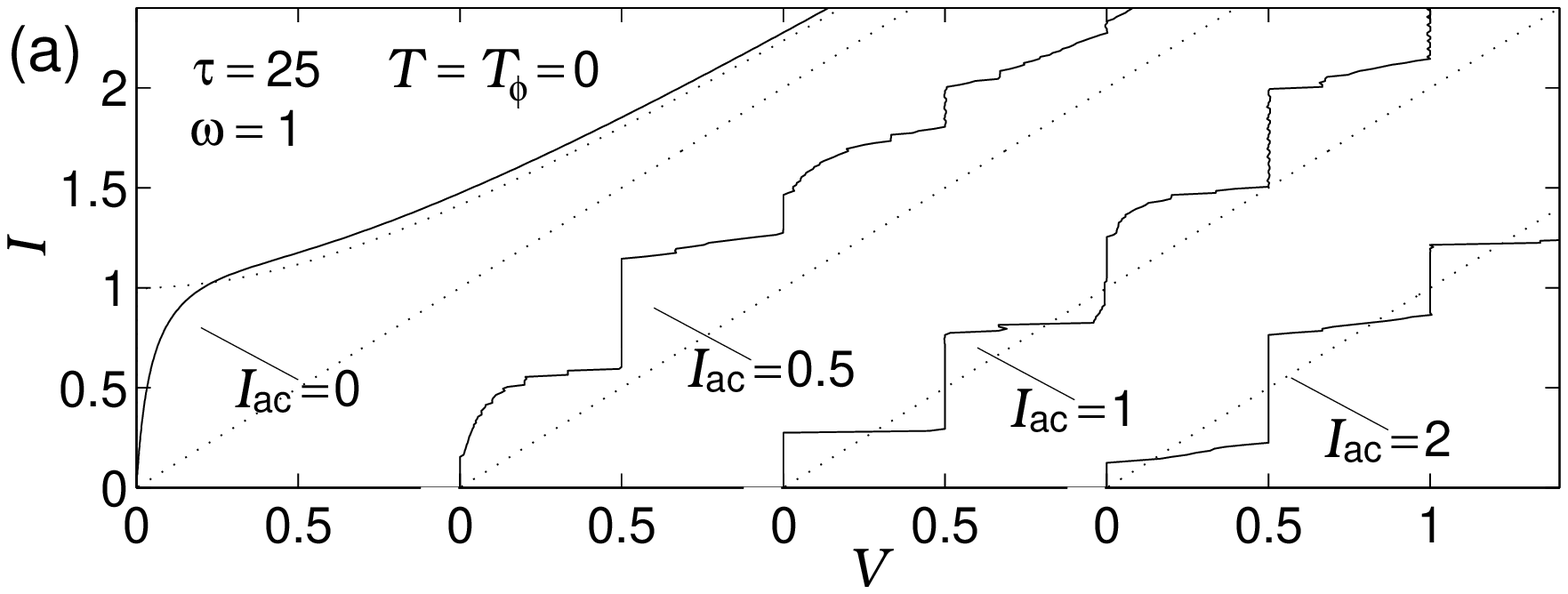}     
\epsfxsize=\hsize 
\epsffile{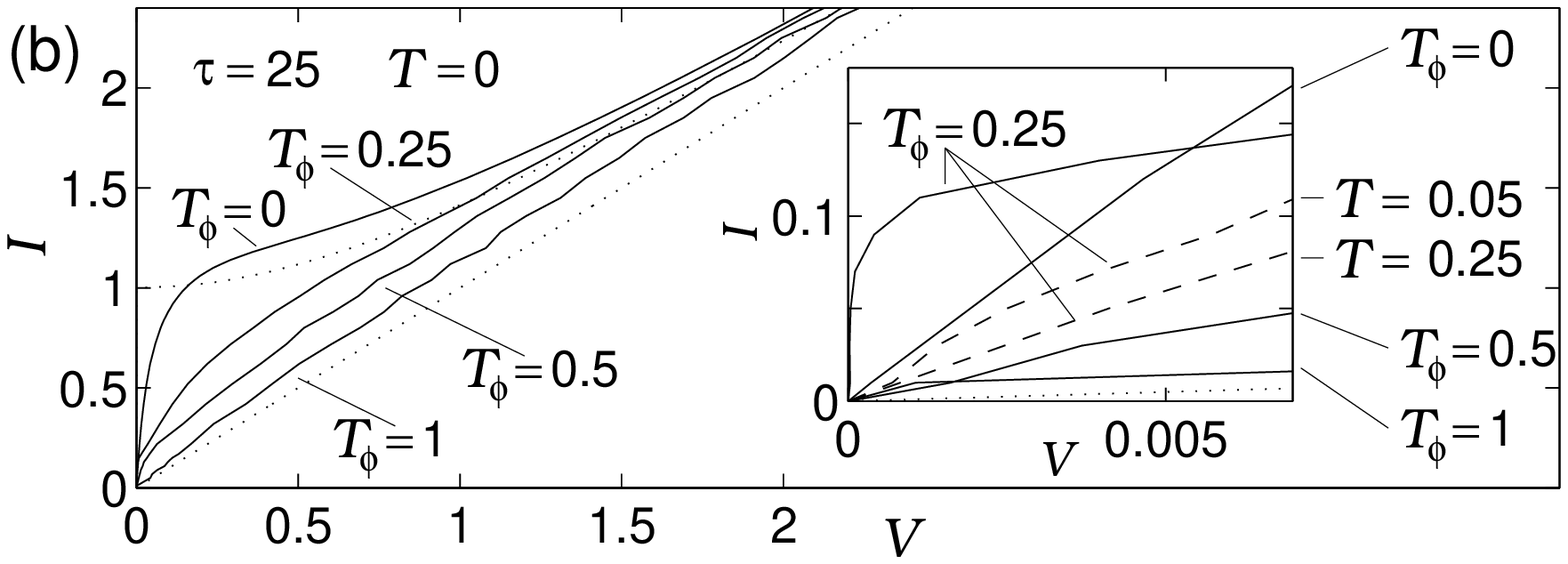}      
\refstepcounter{figure}
\label{fig_shap}
FIG. \ref{fig_shap}:
(a) An added ac current bias (with amplitudes $I_{\rm ac}$, frequency 
$\omega = 2eV_{\rm neq}/\hbar$, and no noise) produces Shapiro steps in the dc $I$---$V$ curves (here $\tau=25$).  The step at $V=0.5$ is due to supercurrents at twice the Josephson frequency.
(b) The effect of external noise (indicated by $T_\phi$) on the dc
$I$---$V$ curves with $\tau=25$ (the waviness is
due to the finite numerical integration time).
Inset: on a much expanded scale the $T_\phi = 0.25$ curve is seen to
cross the one for $T_\phi = 0$, and appears to have a ``critical
current'' of $\sim 0.1 I_{\rm neq}$.
Thermal noise counteracts this ($T > 0$; dashed lines).
\end{figure}

In the presence of a high--frequency external periodic perturbation with frequency $\omega$, the dynamics of the junction may phase--lock with the external drive, producing special features called Shapiro steps at the voltage $V_{\mathrm dc} = \hbar\omega/2e$ or rational multiples thereof.  As in the previous section, the derivation for a voltage--biased situation is simpler and may be done analytically in a variety of limits, but in many experiments a current--biased description is more appropriate \cite{Konrad}.  We therefore take 
$I = I_{\mathrm dc} + I_{\mathrm ac} \cos \omega t$, and integrate \Eq{dervs} numerically.  A sample of the results is given in Fig.~\ref{fig_shap}(a).  The dominance of the second harmonic, 
$\omega = 2(2e/\hbar)V$, is clearly displayed by the strong half--integer Shapiro step at $V=\half (\hbar\omega/2e)$.  Note that when the energy--relaxation is fast, $\omega \tau_{\mathrm E} \ll 1$ and the ac drive is weak, one can apply \Eq{smallV} and Shapiro steps cannot occur --- the ratio of the dc components of the current and the voltage is in this case 
${1 \over 2\pi} \int_0^{2\pi} G(\phi) \, d\phi$ by definition \cite{Jack}.
In practice, because of deviations from the idealized limit, the NEQ Shapiro steps grow linearly with $\omega$ (or $V_{\mathrm dc}$) at small frequencies.  This was indeed observed experimentally in Ref.~\cite{Konrad}.

In order to model noise, one adds fluctuating terms,
$I_{\mathrm F}$ and $J_{\mathrm F}$, to the derivatives of $F$ in \Eq{dervs}, with 
$\aver{ I_{\mathrm F}(t) I_{\mathrm F}(0) } = 2 T_\phi G_{\mathrm N} \, \delta(t)$ and similarly
$\aver{ J_{\mathrm F}(t) J_{\mathrm F}(0) } \propto T \, \delta(t)$.
For thermal noise, $T_\phi=T$; high--frequency external current noise 
gives $T_\phi > T$.  Again, the model may be integrated numerically, and Fig.~\ref{fig_shap}(b) displays some representative results.

\begin{figure}[t]
\epsfxsize=\hsize 
\epsffile{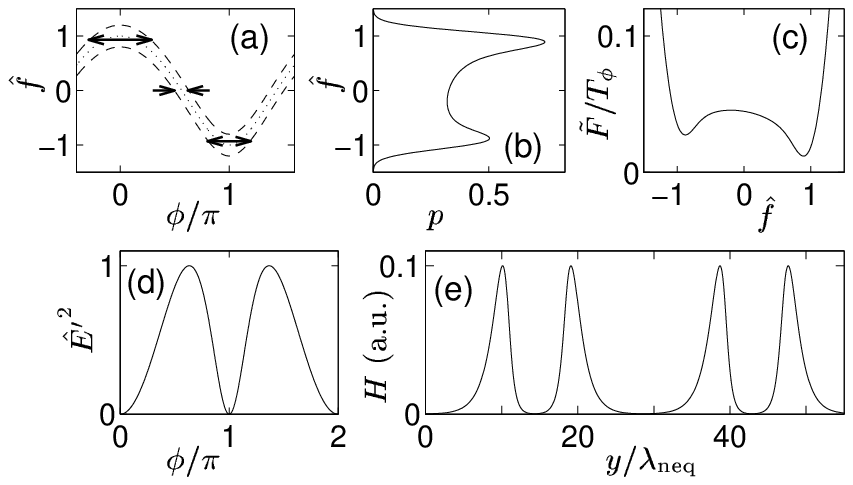}     
\vspace*{1mm}
\refstepcounter{figure}
\label{fig_var}
FIG. \ref{fig_var}:
(a) The ranges of noise--induced $\phi$--fluctuations for fixed 
$\hat f$ (arrows), with contours at $\hat f = \hat E$ (dotted
line) and $F(\phi,\hat f)=0.04E_{\rm neq} = T_\phi$ (dashed lines; $I=0$).
(b) The probability density in $\hat f$, for $\tau_{\mathrm E} \to \infty$,
inferred from (a) is $p(\hat f) \propto \int d \phi \, \exp(-F/T_\phi)$.
(c) The effective potential, $\tilde F(\hat f) = -T_\phi \log p$, has
two dips which may ``trap'' the system.
(d) The effective free energy of a long junction, \Eq{Epsq}.
(e) The resulting spatial variation of an applied magnetic field.
 \end{figure}

The thermal case with $I \to 0$ is particularly simple, because one 
may integrate out dynamical variables.  If $\hat f$ is integrated out, which is also justified for fast relaxation ($\tau_{\mathrm E} \to 0$), then by definition we reproduce the RSJ model, which in the present high--temperature case is trivial because $I_c = 0$.  It is more interesting to take $\tau_{\mathrm E} \to \infty$, and treat $\phi$ as the ``fast variable'', see Fig.~\ref{fig_var}(a).  
For $T_\phi \ll E_{\rm neq}$, the probability density
of $\hat f$ has two peaks [Fig.~\ref{fig_var}(b)], and the 
``effective potential'' $\tilde F(\hat f)$ has valleys of depth 
roughly $\sim T_\phi$ [Fig.~\ref{fig_var}(c)].  
In the case of thermal noise, $T = T_{\phi}$ and this only leads to a 
complicated description of the previous trivial case, but
if $T \ll T_\phi$ the rate of phase--slip is drastically reduced: 
$\hat f$ is attracted to $\pm 1$, and $\phi$ remains near the extrema
of $\hat E$.   This explains the very large slope displayed in the inset of Fig.~\ref{fig_shap} for $T \ll T_\phi \ll E_{\rm neq}$ --- surprisingly, noise can enhance the zero--bias conductance here, as opposed to its effect in the RSJ model.  Upon further expansion of the abcissa, one may observe that the slope never increases beyond $\sim \tau^2$, despite the fact that the present discussion implies that this zero--bias conductance is exponentially large.  This is due to the fact that the fluctuations in $\phi$ induce small fluctuations in $f$ (even when $T=0$), which are of order $1/\tau_{\mathrm E}$ and have been ignored above.

In junctions with a length $L_y$ much longer than the dephasing length 
$\sqrt{D \tau_\varphi}$ (in the experiments of Refs.~\cite{FEL,Bfield,dfrntl,Konrad}, $L_y \sim 100\mu$m), it is reasonable to expect the dynamical variables $\hat f$ and $\phi$ to depend on the position $y$.  In fact, a spatial dependence of $\phi$ may have a prohibitive price in energy, as it is associated with magnetic fields.  To analyze  this situation, the free--energy of \Eq{mod} is generalized to:
\beq \label{Fint}
F[\phi,\hat f]  =   \int_0^{L_y} {\mathrm d}y \; \left[
{\cal F}\leftp \phi,\hat f;{d\phi \over dy},{d\hat f \over dy} \rightp 
- {\hbar \over 2e} {I \over L_y} \phi \right]  \; ,
\eeq
where the gauge--invariant phase difference, $\phi(y)$, and the
local occupation probability of low--lying states, $\hat f(y)$, have become functional variables, and
\beq \label{abc}
{\cal F}(\phi,\hat f;\phi',\hat f')  =  
\alpha\left( \hat f - \hat E(\phi) \right)^2 + 
\beta {\phi'}^2 + \gamma \hat {f'}^2   \; . \label{Gsum}
\eeq
\Eq{dervs} is then replaced by 
\beq \label{fdervs}
{\hbar \over 2e} {\partial\phi \over \partial t}  = 
   - {L_y \over G_{\mathrm N}} \, {2e \over \hbar} {\delta F \over \delta \phi} 
\quad ; \quad
{\partial \hat f \over \partial t}  =  
- {L_y \over 2\tau_{\mathrm E} E_{\rm neq}} 
{\delta F \over \delta \hat f}  \; .
\eeq
Here $\alpha = E_{\rm neq}/L_y$, 
$\beta = (1/8\pi)(\hbar c/2e)^2(L_z/L_x)$ is the energy--density
of the magnetic fields associated with the spatial dependence of the gauge--invariant phase, $H \propto (\partial\phi/\partial y)$ (Ref.~\cite{geom}), and $\gamma = \alpha \tau_{\mathrm E} D_y$ is introduced to give diffusion of $\hat f$ in the $y$ direction.  For simplicity, we set \cite{Dy} the coefficient $D_y$ equal to $D$.

We now turn our attention to some interesting effects generated by thermal noise in long junctions.  Fluctuations in the occupations, i.e.\ the space--dependent generalization of $\Delta f \s\sim N^{-1/2}$, produce fluctuations in the supercurrents $\delta F/\delta \phi$ (as in the case of Johnson--Nyquist noise, these are related by a fluctuation--dissipation theorem to the zero--bias conductance, which in turn is enhanced as discussed in the previous section \cite{fluc-diss}).  The magnetic fields generated by these fluctuating supercurrents, and the accompanying 
$y$--dependent fluctuations of $\phi$, generate a $y$--dependence also in 
$\hat E$.  As a result, $\hat f$ relaxes to a value of $\hat E$ which is spatially averaged over a length--scale $\sqrt{D \tau_{\mathrm E}}$, and can deviate from its local value, giving rise to a finite energy density. 
Indeed, after rewriting the $y$ integral in \Eq{Fint} as a sum over wavenumbers, $L_y \sum_q$, one may integrate out $\hat f$ one wavenumber at a time, which leads to a replacement of the first and last terms of \Eq{abc} by 
$\alpha \LP 1-(1+D \tau_{\mathrm E} q^2)^{-1} \RP |\hat E_q|^2$,  (this is seen by completing the square in the exponent).  The next step is to assume that $\beta \gg T \sqrt{D \tau_{\mathrm E}}$ (the experimentally relevant regime), and to introduce a cutoff $\Lambda$ such that $T/\beta \ll \Lambda \ll (D \tau_{\mathrm E})^{-1/2}$.  The fluctuations of $\phi$ over regions $< \Lambda^{-1}$, are small because of the large value of $\beta$ assumed, and one may therefore linearize $\hat E(\phi)$ within each such region, i.e.\ take 
$\hat E_q \simeq (d\hat E/d\phi) \phi_q$ and replace $L_y \sum_q$ by 
$(1/2\pi) \int {\mathrm d}y \, {\mathrm d}q$.  One may now integrate over the components of $\phi$ with wavenumbers $> \Lambda$, taking care to keep and re--exponentiate the coefficients $c$ in the integrals of the type $\int {\mathrm d}\phi_q \, \exp(-c|\phi_q|^2)$, as they now contain an essential $\phi$--dependence.  For $\gamma \ll \beta$, the resulting free energy is given, up to a constant, by \cite{accu}
\beq \label{Epsq}
\tilde {\cal F}(\phi,\phi')  \l=  
{T \over 2 \sqrt{D \tau_{\mathrm E}}}
{\gamma \over 2\beta} \leftp {d\hat E \over d\phi} \rightp^2 + 
\beta {\phi'}^2  \; ,
\eeq
where $\tilde {\cal F}$ is defined as in \Eq{Fint}, and $\phi$ is assumed smooth on the scale $\Lambda^{-1}$.
The system thus has a non--vanishing energy density even when 
$\phi' = 0$, and differs from a conventional long Josephson 
junction only in the replacement of $- \cos \phi$ in the usual 
description by a $\phi$--dependence with two equal--valued minima per period, see Fig.~\ref{fig_var}(d).

In this model, an external magnetic field ($H_z$) produces
Josephson--like vortices containing half a flux quantum each,
$hc/4e$, see Fig.~\ref{fig_var}(e).  This may explain the magnetic
period--halving effects of Ref.~\cite{Bfield}.  The ``Josephson--like
penetration depth'', defined by $\lambda_{\rm neq}^2 =
2 \sqrt{D \tau_{\mathrm E}} \beta^2/\gamma T$ (Ref.~\cite{accu2}), is in between the exponential
decay--length of the magnetic field in regions with $\phi \simeq 0$
(mod $2\pi$) and the smaller one for $\phi \simeq \pi$ regions.  The latter two may be obtained by multiplying $\lambda_{\rm neq}^2$ by
$(\half \hat {E'}^2)'' \neq 1$, where the primes denote differentiation with respect to $\phi$.  True period--halving occurs
only when $\hat E \simeq \cos \phi$, e.g.\ when 
$\hbar/\tau_\varphi > E_{\mathrm C}$, and NEQ effects are weak.

If $L_y \ll \lambda_{\rm neq}$, then $\phi(y) \simeq {\rm const}$.
The junction prefers both $\phi \simeq 0$ and $\phi \simeq \pi$
values, with intervening energy barriers of
$2\pi (2e/\hbar c)^2 \sqrt{D \tau_{\mathrm E}} \, T E_{\rm neq} (L_x/L_z)$ [This is just the coefficient in \Eq{Epsq} multiplied by $L_y$ --- the maximal value of $\hat E'$ is here equal to 1].
In dc measurements which are not phase--sensitive, this behaves as a Josephson coupling energy, which persists to high temperatures, 
$T \gg E_{\mathrm C}$, and is roughly 
$\sim (D / 10 \, cL_z) E_{\mathrm J}(0)$ (Ref.~\cite{geom}).  This estimate is obtained by assuming that $\sqrt{D\tau_{\mathrm E}}$ is equal to a few times $L_x$.  If indeed $\sqrt{\tau_{\mathrm E}}$ does not depend strongly on temperature, one obtains a possible mechanism for the unexplained coupling energy observed in the experiments of Ref.~\cite{dfrntl}.

The behavior of $\phi(y)$ in this last case is analogous to that 
of a long polymer diffusing in a disordered gel \cite{Edwards}: the polymer may be pinned by
``attractive'' disordered regions, leading to exponential suppression of diffusion --- whenever one part of the polymer begins diffusing away
from the pinning center, it is pulled back by forces due to the thermal
motion of its other parts.  If the polymer were replaced by point--like monomers, the diffusion of its constituents would not be hindered, and the ``attractive'' nature of such regions would not be noticeable.  Likewise, if the junction of length $L_y \ll \lambda_{\rm neq}$ is replaced by much shorter junctions of length 
$< \sqrt{D \tau_{\mathrm E}}$, the high--temperature Josephson coupling discussed here would disappear, and only the enhanced conductance discussed in the previous section (see Fig.~\ref{fig_IV}) would remain.

\section{Clean junctions}

\begin{figure}[t]
\epsfxsize=\hsize
\epsffile{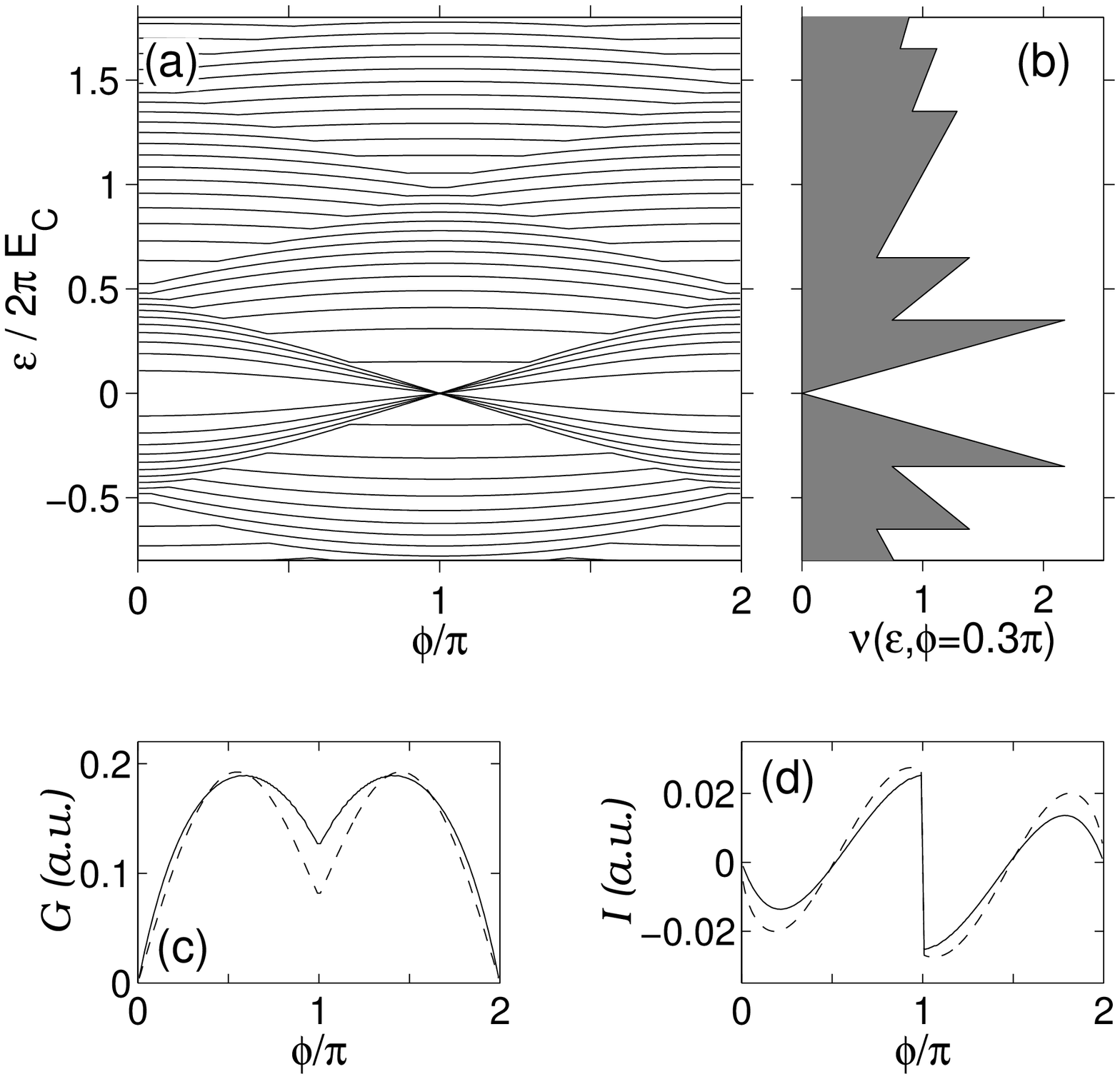}
\refstepcounter{figure}
\label{fig_clean}
FIG. \ref{fig_clean}:
(a) Energies $E_n(\phi)$ of representative Andreev bound states, for a (three--dimensional) quasiballistic S--N--S junction, as in Fig.~\ref{fig_dos}.  Here ``quasiballistic'' means that there is enough disorder to produce avoided crossings (not resolved), but not enough to modify the density of states appreciably from that of the super--clean limit.
(b) The density of states of the junction, sampled at $\phi=0.3\pi$.
The NEQ supercucrrents in the limits of small and large 
$(2e/\hbar)\tau_{\mathrm E} V$ are plotted in (c) and (d) respectively.  The model using a simple representative curve (dashed lines) is compared to the result of the full sum over the different $E_n(\phi)$ levels (full lines).  
\end{figure}

The starting point for the theory developed here was the fact that the spectrum of an S--N--S junction depends on the phase $\phi$, but only the case of a diffusive junction was made explicit.  Motivated by the experiments \cite{FEL,Bfield,dfrntl,Konrad}, we will next discuss a specific clean (or qusiballistic) system, with 
$N_\perp L_x \gg l_{\mathrm el} \gg L_x$.  Consider first the density of states for a rectangular geometry in the ideally clean case.  The spectrum may be obtained for each transverse channel separately, by using Bohr--Sommerfeld quantization and introducing the $\phi$--dependence through the phase--shifts associated with the Andreev reflections at the two N--S interfaces \cite{Sipr}.  This gives
\beq \label{cleans}
\nu(\epsilon,\phi) = 
\sum_{\pm\phi} \sum_{i=1}^{N_\perp} \sum_{m=-\infty}^\infty
\delta \LP \epsilon - {\hbar v_i\over 2L_x} 
\LP (2m-1)\pi \pm \phi \RP \RP  \; ,
\eeq
where $v_i$ is the component of the Fermi velocity in the $x$ direction for the $i$th transverse channel.  We will assume for simplicity that the junction is three--dimensional, i.e.\ that there are two substantial transverse directions.  The sum over transverse channels, $\sum_{i=1}^{N_\perp} \dots$, may then be replaced by the integration $N_\perp \int_0^{v_{\mathrm F}} \dots 
2v_\perp \, {\mathrm d}v_\perp / v_{\mathrm F}^2$, where 
$v_i = \sqrt{v_{\mathrm F}^2 - v_\perp^2}$.  This has the effect of replacing every $\delta$--function in \Eq{cleans} by a ``sawtooth'':
\beq \label{teeth}
\nu(\epsilon,\phi) =  N_\perp \sum_{\pm} \sum_m
{2|\epsilon| \over \epsilon_{m,\pm}^2} 
\Theta \LP \epsilon_{m,\pm} - |\epsilon| \RP  \; ,
\eeq
where $\epsilon_{m,\pm} = \half E_{\mathrm C} 
\LP (2m-1)\pi \pm \phi \RP$, the Thouless energy is 
$E_{\mathrm C} = \hbar v_{\mathrm F}/L_x$, and $\Theta$ is the Heavyside step--function.  This density of states rises linearly at small energies \cite{minigap}, see Fig.~\ref{fig_clean}, as opposed to the mini--gap of the diffusive case 
[for a two--dimensional junction, each term in the sum would be divided by 
$2 \sqrt{1-\epsilon/\epsilon_{m,\pm}}$, introducing divergencies in $\nu(\epsilon,\phi)$; as these divergencies are integrable, the NEQ results would still be qualitatively similar].

Weak elastic scattering will result in avoided crossings in the spectrum.  The disorder--averaged excitation energy of the $n$th state may then be introduced as above, $n = \int_0^{E_n} {\mathrm d}\epsilon \, \nu(\epsilon,\phi)$.  These are displayed in Fig.~\ref{fig_clean}, which is to be compared with the diffusive case of Fig.~\ref{fig_dos}.  We are assuming here that $\hbar/\tau_{\mathrm el} \ll E_{\mathrm C}$ and therefore the elastic scattering does not appreciably smear the features in $\nu(\epsilon,\phi)$.  Note that the parameter $\tau_{\mathrm el}$ is nominal and describes the scattering rate in the absence of proximity effects; the proximity effects change the scattering rates, which are given by the product of $\nu(\epsilon,\phi)$ with the variance of the elastic--scattering matrix--elements.  Thus the typical change in $\epsilon$ suffered by an individual level due to the disorder may be substantially increased or decreased relative to 
$\hbar/\tau_{\mathrm el}$.  This is particularly relevant in the region near $\phi=\pi$ and $\epsilon=0$, where all of the $N_\perp$ states associated with the lowest--lying term in \Eq{teeth} become crammed into a very small range of energy.  We will ignore such complications here --- their resolution (e.g.\ the size of the region in $\epsilon$ and $\phi$ which is affected) depends on the value of 
$E_{\mathrm C} \tau_{\mathrm el} / \hbar$ which we are taking to be infinite.  In fact, one should also note that apart from assuming $\Delta \gg \epsilon$, the Andreev approximation used in deriving \Eq{cleans} is only valid when the kinetic enery in the $x$ direction, $\half m^* v_i^2$, is much larger than the gap $\Delta$ (here $m^*$ is the effective mass of the electrons).  In practice, a fraction 
$\Delta/E_{\mathrm F}$ of the states do not have a large enough $v_i$; although they are not strongly affected by Andreev reflections (``glancing angle'') and show much weaker proximity effects \cite{Sipr}, they are ignored here by taking $E_{\mathrm F}/\Delta$ to infinity.
 
We next consider dynamic NEQ effects in this system.  For brevity, only the results in the simple limits of small and large voltages relative to the inelastic scatterring rate, Eqs.~(\ref{smallV}) and (\ref{largeV}), are plotted in Fig.~\ref{fig_clean}(c) and (d).  Note that at moderatley large voltages, $(2e/\hbar)V \sim 1/\tau_{\mathrm el}$, the avoided crossings may ``no longer be avoided'', and the ballistic structure of the spectrum may substantially alter the analysis.  In other words, the occupations might, under sufficiently strong drive, become dependent not only on energy but also on the angle of propagation relative to the $x$ direction (i.e.\ the transverse channel index $i$).  As explained in Sec.~II, the use of the relaxation--time approximation, \Eq{relax}, is also expected to fail at large voltages, when $(2e/\hbar) V  \stackrel{>}{\sim}  \sqrt{E_{\mathrm C} /\hbar \tau_{\mathrm E}}$.   Either one of these restrictions may be more stringent for a particular application, because $\tau_{\mathrm E}$ is usually much larger than $\tau_{\mathrm el}$, which itslef is $\gg \hbar/E_{\mathrm C}$ in the clean case.

As in the case of the diffusive junction, it is proposed here that the complicated energy--dependence of the spectrum can be ignored to a first approximation, with the phase--dependence and the NEQ occupation followed only for a single representative level.  The individual curves $E_n(\phi)$ in Fig.~\ref{fig_clean} have discontinuous derivatives (because $\nu$ is discontinuous), and are not represented well by any single one of them.  Instead we choose the form
$\hat E(\phi) = 1-2\sqrt{|\sin(\half\phi)|-\half|\sin(\phi)|}$, which reproduces the qualitative behavior of $G(\phi)$ near both $\phi=0$ (where it vanishes linearly) and $\phi=\pi$ (where it approaches a constant).  This form has no adjustable parameters, and gives errors of order 30\%, as shown in Fig.~\ref{fig_ac} [the value of $E_{\mathrm neq}$ used was 
$2.0 (E_{\mathrm C}^3/\delta T)$].  These errors could of course be reduced by using a more complicated form of $\hat E(\phi)$, but it turns out that the contribution of the levels near 
$\epsilon = \pi E_{\mathrm C}$ end $\phi = 0$ is difficult to model with a single representative curve (to correct this, one could perhaps introduce a $\phi$--dependent number of represented levels $N$, which would vanish linearly at small $\phi$).  Despite these differences in $\hat E(\phi)$, and even the possible need to follow the occupations of many levels instead of using a single representative, we expect that the results plotted in Figs.~\ref{fig_IV} and \ref{fig_shap} would not be dramatically altered if the calculation were redone for a clean junction.  The results for long junctions, Fig.~\ref{fig_var}, would obviously be more sensitive to the qualitative behavior near $\phi=0$ and $\phi=\pi$.  However, for any realistic system there will be additional effects which would ``soften'' this behavior (see below).

\section{Discussion}

In summary, it was shown here that in wide S--N--S junctions, significant deviations of the electronic occupations from equilibrium may be caused not only by an external dc or ac drive, but also by external noise, or even by space--dependent thermal noise in the case of long junctions.  Such non--equilibrium occupations have structure in the excitation energy $\epsilon$ on the scale of the Thouless energy $E_{\mathrm C}$, and therefore have ``ensemble coherence lengths'' large enough to span the width of the junction $L_x$.  For this reason, even when the temperature $T$ is high, and the equilibrium ``ensemble coherence length'', $\xi_{\mathrm N}$, is small enough to give an exponential supression of the equilibrium Josephson effect, supercurrents may flow in the junction.  These supercurrents lead to Josephson--like characteristics of the junction, and are charactrized by an approximate period--halving and a weak temperature dependence.

\begin{figure}[t]
\epsfxsize=\hsize
\epsffile{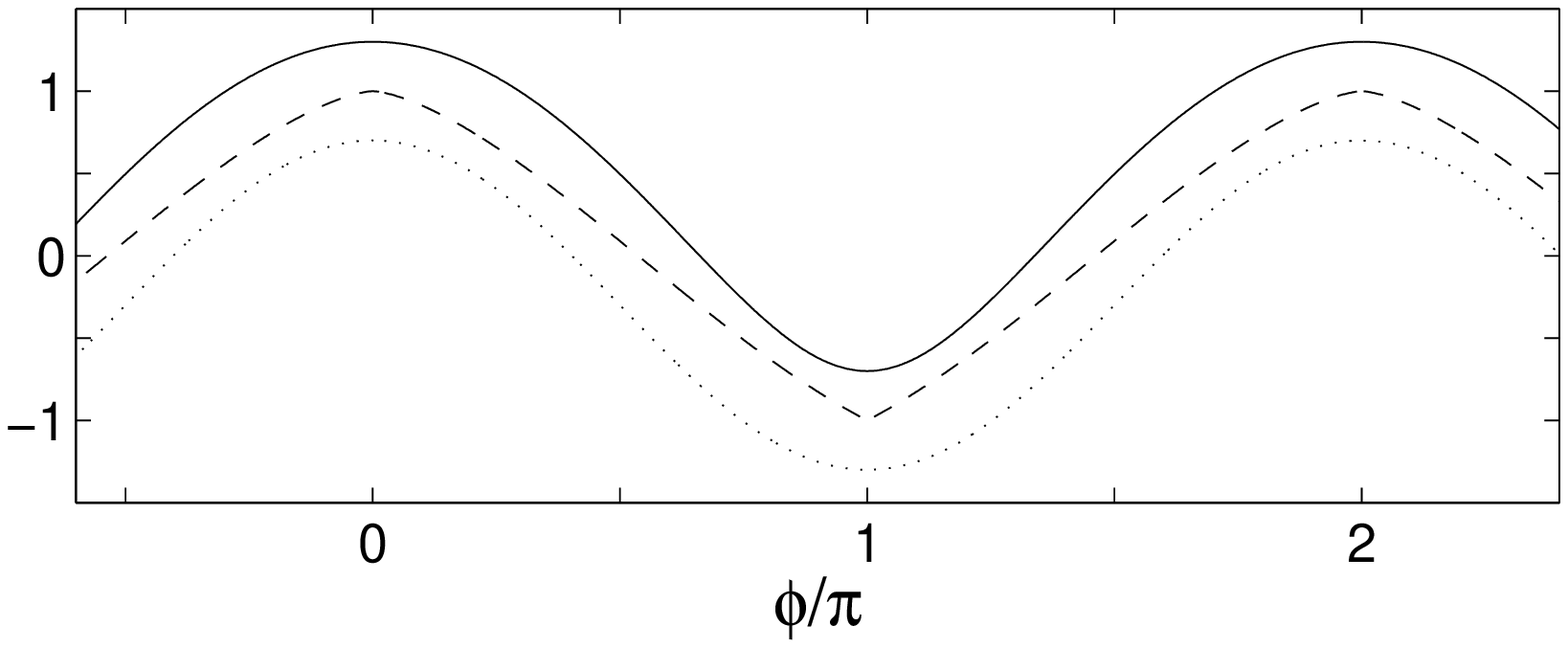}
\refstepcounter{figure}
\label{fig_shapes}
FIG. \ref{fig_shapes}:
Sketch of the different versions of $\hat E(\phi)$ suggested for a diffusive junction (full line, Sec.~II), a clean junction (dashed line, Sec IV), and the simple cosine used in modelling the experiment (dotted line, Ref.~\cite{dfrntl}), shifted for clarity.  The ``bottom of the valley'' shown in the inset of Fig.~\ref{fig_IV} and in Fig.~\ref{fig_var}(a) should be modified accordingly.
\end{figure}

The model developed here to describe these effects, defined by \Eq{mod} and \Eq{dervs}, has been compared with experimental results in Ref.~\cite{Konrad}, for both dc and ac effects, [Fig.~\ref{fig_IV} and Fig.~\ref{fig_shap}(a) respectively].  It was found there that the high harmonics are very weak, and the model with $\hat E = \cos\phi$ (see Fig.~\ref{fig_shapes}) describres both types of effects well with the same choice of parameters, $I_{\mathrm neq}$ and $\tau_{\mathrm E}$ (the latter was further corroborated by estimating it from the sub--gap structure).  This may serve as additional motivation for representing all of the states in the spectrum by a single curve $\hat E(\phi)$, as was done here --- the fine details which are being left out by this procedure are difficult to observe experimentally.  The most important factor which contributes to the ``softening'' of the phase--dependence, compared to the clean system ($l_{\mathrm el} \gg L_x$) considered in the previous section, is probably the elastic scattering, $l_{\mathrm el} \sim L_x$, in the experiment.

Theoretically, there are many refinements of the model which remain to be pursued, of which three are listed here: (a) No distinction has been made here between electron--electron and electron--phonon inelastic scattering; instead all mechanisms which smooth the occupation functions on the scale of $E_{\mathrm C}$ have been lumped into one parameter, 
$\tau_{\mathrm E}$.  Although this may be adequate, the relationship of this parameter to $\tau_{\mathrm e-e}$ and $\tau_{\mathrm e-ph}$ needs to be examined more carefully, especially as it will determine its temperature dependence.  (b) The low--voltage limitation of the present model, $eV \ll \sqrt{E_{\mathrm C} \hbar / \tau_E}$, may perhaps be relaxed to $eV \ll E_{\mathrm C}$ without significantly complicating the model, by adding a phenomelogical term involving 
$D_{\mathrm E} \propto V^2$ to the evolution equation of $\hat f$, \Eq{relax}.  This is important because experimental $I$--$V$ curves often exhibit interesting features in this range of voltages \cite{Konrad}.
(c) The lateral diffusion of $\hat f$ in long junctions has been treated here in a phenomelogical manner, by setting $D_y = D$; a more sophisticated treatment of this may be in order (cf.\ Ref.~\cite{Dy}).

Clearly there is a richness of phenomena in such junctions which 
remains to be explored experimentally as well.  The most dramatic challenge, perhaps, is to observe the predicted ``half flux quanta'' of
Fig.~\ref{fig_var}(e) directly, e.g.\ by pinning them and using
magnetic microscopy techniques similar to those used in observing half flux quanta in junctions made of high--$T_{\mathrm c}$ materials \cite{Kirtley}.

I thank S.J. Allen, D.V. Averin, S.F. Edwards, M.P.A. Fisher, J.G.E. Harris, H. Kroemer, K.W. Lehnert, J.P. Sethna, B. Spivak, M. Tinkham, and F. Zhou for fruitful discussions.  This work was supported by NSF grants No.\ PHY94-07194 and No.\ DMR96-30452.

\section*{Appendix: Comparison with the Green's functions approach}

In order to compare our results with those obtained with the Green's function approach in the literature, we first quote the Usadel equations, in the form used by Zhou {\it et al.} \cite{minigap}.  In the normal--metal part of the diffusive junction of Fig.~\ref{fig_dos}, the equations read:
\beqa \label{Usad}
{D \over 2} \theta'' + i \epsilon \sin \theta - 
{D \over 4} (\chi')^2 \sin 2\theta  & = &  0   \; ;  \cr
(\chi' \sin^2 \theta)' & = & 0    \; ,
\eeqa
in terms of the variables $\theta$ and $\chi$, which are both complex and depend on position $x$ and excitation energy $\epsilon$ (the primes denote differentiation with respect to $x$).  The excitation energy $\epsilon$ here is taken to be positive (at $\epsilon=0$ the variables $\chi$ and $\theta$ are real and can be thought of as angles).  Here $\hbar$ has been set to 1, and $D$ is the diffusion constant as before.  The boundary conditions are $\theta = \half\pi$ and 
$\chi = \pm \half \phi$ at the N--S interfaces, $x = \pm \half L_x$.  For a recent derivation and numerical implemetation of these equations, see Ref.~\cite{Argaman}.

Once the Usadel equations have been solved, one has the Green's functions (in angular variables), and may obtain many of the properties of the junction.  Specifically, the local density of states is given by 
$\real \cos \theta$ times its value in the absence of proximity effects, and the Josephson current density, $j(\epsilon,\phi)$, is given by the imaginary part of $\chi' \sin^2 \theta$.  The fact that this must be equivalent to the mesoscopic--systems approach used in \Eq{Isum}, i.e. that $j(\epsilon,\phi) = (2e/\hbar) \nu (dE_n/d\phi)$
is clear if one recalls that the phase $\phi$ can be represented through a vector potential, and that a derivative of the energy operator with respect to the vector potential gives the current operator.
It is a non--trivial excercise to derive this relationship analytically from the disorder--averaged \Eq{Usad}, but it is easy to check it in practice once the solutions have been obtained, and this has in fact been applied to evaluate the accuracy of the numerical scheme used to plot Fig.~\ref{fig_dos}.  Note also that the normal current $I_{\mathrm N}$ appears in the Green's function approach, e.g.\ in Ref.~\cite{Lempitskii}, in exactly the same way as in \Eq{Isum}.

We next turn to the evolution of the occupations $f_n$.  In order to represent both electrons and holes without extending $\epsilon$ to negative energies, the Green's--functions approach uses two occupation functions: the symmetric combination 
$f_1(\epsilon,x) = 1 - f(-\epsilon,x) - f(\epsilon,x)$, and the antisymmetric combination
$f_2(\epsilon,x) = f(-\epsilon,x) - f(\epsilon,x)$ [note that the sum over $n$ in \Eq{Isum} may be restricted to positive energies by replacing $f_n$ by $-f_2$, because $E_{-n}=-E_n$].  In equilibrium, one has $f_2 = f_0 = \tanh(\epsilon/2T)$ and 
$f_1 = -e\varphi \, \partial f_2 / \partial \epsilon$, where a
small scalar potential $\varphi(x)$ has been introduced.  In the present work, we have essentially introduced the electric field through a time--dependent vector potential, so that $f_1 = 0$ throughout (cf.\ Ref.~\cite{imbal}).  The evolution of $f_2$, which is space--independent for $eV \ll E_{\mathrm C}$, is controlled by
\beq \label{kin}
{\partial f_2 \over \partial t} + {\partial f_2 \over \partial \epsilon} \, {dE_n \over d\phi} \, {d\phi \over dt} = 
-{1 \over \tau_{\mathrm E}} (f_2-f_0)  \; .
\eeq
This is just \Eq{relax} rewritten using the variables $\epsilon$ and $t$ rather than $n$ and $t$ as independent (we have $f_2 = 1-2f_n$ when $\epsilon = E_n$).  By noting that $(dE_n/d\phi)(d\phi/dt) = (j/\nu)V$, one finds that \Eq{kin} is identical to Eq.~(12) of Ref.~\cite{Lempitskii}, where a scalar potential with 
$\varphi = (x/L_x) V$ has been used instead of a vector potential, and therefore the difference between the values of $f_1$ evaluated at the two N--S interfaces is equal to $eV \partial f_2 / \partial \epsilon$ (Ref.~\cite{error}).

We have thus shown that Eqs.~(\ref{Isum}) and (\ref{relax}), which form the basis for the present work, are equivalent to those underlying more conventional treatments of NEQ effects.


\begin{references}
\bibitem{sample} Note that the above list of references is only a small sample of the relevant literature.
\bibitem{condens} The interaction energy or condensation energy of the superonducting banks is not included in $\sum_n E_n f_n$, but as we are considering the high--temperature regime, $\xi_{\mathrm N} \ll L_x$, it is $\phi$--independent.
\bibitem{imbal} Here $f_{-n} = 1-f_n$; there is no ``charge imbalance''
\cite{Tinkham}.
\bibitem{strict} $E_{\rm rep}(\phi)$ is defined by
$N f_{\rm eq}(E_{\rm rep}) = \sum_{n=1}^N f_{\rm eq}(E_n)$; thus
$F(\phi,\hat f) \propto 
\log( {\rm weighted~\#~of~configs.~with~the~given~} f)$.
\bibitem{geom} $L_x$ and $L_z$ should be adjusted for magnetic 
``leakage'', e.g. $L_x \to L_x+2\lambda$, with $\lambda$ the
penetration depth in S.
\bibitem{Dy} However, diffusion among low--lying states may be
modified by proximity effects in a $\phi$--dependent manner.
\bibitem{accu} The term obtained in the exponent is 
$- (1/2\pi) \int_0^{L_y} {\mathrm d}y \int_\Lambda^\infty {\mathrm d}q \, 
\log \LP 1 + 2X/(1+D\tau_{\mathrm E}q^2) \RP$, where 
$X=(\gamma/2\beta) \hat E'^2$.  For $\gamma \ll \beta$ this gives \Eq{Epsq}; otherwise $\sqrt{1 \s+ 2X} - 1$ replaces $X$ there, and $\beta$ is replaced by $\beta + \gamma$.
\bibitem{accu2} It is possible that in the presence of an applied magnetic field, it will be necessary to adjust the cutoff $\Lambda$ introduced above, so as to make it $\gg \lambda_{\mathrm neq}^{-1}$.
\bibitem{error} However, Ref.~\cite{Lempitskii} later replaces 
$\partial f_2 /\partial\epsilon$ in \Eq{kin} by 
$\partial f_0/\partial\epsilon$, because thge values of $f_1$ are evaluated inside the superconductors which are assumed to be in equilibrium, rather than at the N--S interfaces.  As the density of states decays rapidly into the superconductors for $\epsilon < \Delta$, all the quasiparticles are (Andreev) reflected back into the normal--metal region before they can equilibrate; thus the equilibrium value of $f_1$ in the superconductor is meaningless.  This replacement is nevertheless justifiable in near--equilibrium situations, 
$(2e/\hbar) V \tau_{\mathrm E} \ll 1$, where $f_2 \simeq f_0$. It has the effect of making it appear to be unnecessary to switch from the $\epsilon$ variable to the $n$ variable.  In the opposite limit, $\tau_E \to \infty$ (i.e.\ assuming slow relaxation even inside the superconductor), the error becomes obvious as there are no relaxation mechanisms and the dynamics of the junction cannot depend on $f_0$.  We note in this context that earlier in Ref.~\cite{Lempitskii} the voltage V was treated as a small parameter with all equations expanded to first order in it; this already introduced a dependence on $f_0$ even when $\tau_{\mathrm E} \to \infty$, and also precluded a discussion of the energy diffusion ($D_{\mathrm E}$) term in the kinetic equations.

For an apparently unrelated reason, Ref.~\cite{Spivak} also does not make the distinction between following the occupations at a fixed $\epsilon$ vs.\ following them at a fixed $n$.  In this case, only the dissipative  dc component of the current was discussed, and presented as an example of the Debye relaxation mechanism.  The expressions for the Debye mechanism used there were originally developed for systems which do not have superconducting elements and thus have much smaller phase--induced variations in $E_n$, of order $\delta$ rather than $E_{\mathrm C}$.  On the other hand, it is clear from Refs.~\cite{Tinkham,Averin,Bagwell} that the importance of following the occupations of the levels in $n$ rather than $\epsilon$ is well understood.  In any case, correcting this does not qualitatively alter the results reported in Refs.~\cite{Lempitskii} and \cite{Spivak}.
\end{references}
\end{document}